\documentclass{JHEP3}
\usepackage[english]{babel}
\usepackage{amssymb}
\usepackage{amsfonts}
\usepackage{amsmath}
\usepackage{graphicx}
\usepackage{epsfig}
\usepackage{cite}
\newcommand{\cE}{{\cal E}}

  \newcommand{\cN}{{\cal N}}
  
  \newcommand{\cR}{{\cal R}}
  \newcommand{\cT}{{\cal T}}

\newcommand{\bR}{{\mathbf R}}

\def\wn{{\mathfrak{w}}}  \def\qn{{\mathfrak{q}}}

\newcommand{\be}{\begin{equation}} \newcommand{\ee}{\end{equation}}
\newcommand{\bea}{\begin{eqnarray}} \newcommand{\eea}{\end{eqnarray}}
\newcommand{\beann}{\begin{eqnarray*}}  \newcommand{\eeann}{\end{eqnarray*}}
\newcommand{\bfig}{\begin{figure}} \newcommand{\efig}{\end{figure}}
\newcommand{\ba}{\begin{array}} \newcommand{\ea}{\end{array}}
\newcommand{\bcen}{\begin{center}} \newcommand{\ecen}{\end{center}}
\newcommand{\btab}{\begin{tabular}} \newcommand{\etab}{\end{tabular}}

     \def\sign{\operatorname{sign}}
\renewcommand{\Re}{\mathop{\rm Re}}   \renewcommand{\Im}{\mathop{\rm Im}}
\def\wn{{\mathfrak{w}}}  \def\qn{{\mathfrak{q}}}
\newtheorem{Proposition}{Proposition}[section]

\newtheorem{Theorem}{Theorem}[section]
\newtheorem{Lemma}{Lemma}[section]
\newtheorem{Corrolary}{Corrolary}[section]

\newcommand{\bp}{\begin{Proposition}}   \newcommand{\ep}{\end{Proposition}}
\newcommand{\bt}{\begin{Theorem}}   \newcommand{\et}{\end{Theorem}}
\newcommand{\bl}{\begin{Lemma}}     \newcommand{\el}{\end{Lemma}}
\newcommand{\bc}{\begin{Corrolary}} \newcommand{\ec}{\end{Corrolary}}
\title{Stability of linear dilaton black holes at the Hagedorn temperature}
\author{Gaetano Bertoldi,${}^a$ Carlos Hoyos,${}^b$ \\
  ${}^a$University of Toronto\\
  ~\,60 St.George St.\\
  ~\,Toronto M5S 1A7, Canada\\
  ~\,E-mail: \email{bertoldi@physics.utoronto.ca}\\
  ${}^b$Department of Physics,
   University of Washington\\
  ~\,Seattle, WA 98915-1560, USA\\
  ~\,E-mail: \email{choyos@phys.washington.edu}}
 \abstract{
We study the thermodynamics and the small fluctuations of a linear dilaton black hole, in an S-dual version of the near-horizon limit of type IIB NS5 branes. The thermodynamical analysis shows that the black hole is in a Hagedorn phase, with marginal stability. The dynamical analysis confirms that the speed of sound of the dual theory vanishes and that there are no evident instabilities at the level of supergravity. We clarify the physical meaning of some singularities of the retarded correlator that were thought to lead to instabilities.
 }


\begin{document}

\section{Introduction}

Little string theories (LSTs) are non-local theories without gravity that describe the dynamics of NS5 branes in the limit where the string coupling vanishes \cite{Seiberg:1997zk, Aharony:1997an, Aharony:1999ks}. Although the excitations on the brane decouple from the bulk, the theory on the NS5 branes remains strongly coupled. LST admits a holographic description in terms of a near-horizon linear dilaton geometry \cite{Aharony:1998ub}. At finite temperature, the only geometries known analytically correspond to a Hagedorn phase, so the temperature is fixed at the scale of the LST.

There are several features that make these theories interesting.
The low energy description of LST in type IIA is given by the (2,0) superconformal theory, whereas for  LST in type IIB
the low energy description is in terms of Super Yang Mills theory with (1,1) supersymmetry.
The holographic description of LST generalizes the more usual AdS/CFT correspondence \cite{firstadscft}. However, the theory has typical properties of a string theory like T-duality,
and the holographic dual allows to compute off-shell quantities. This is in contrast with critical string theories, where so far it is possible only to calculate the S-matrix. Another interesting point is that the linear dilaton black holes are closer relatives to black holes in asymptotically flat space \cite{Minwalla:1999xi,Peet:1998wn,Marolf:2006bk}. Then, understanding Hagedorn LSTs may be a step towards the implementation of holographic ideas to less exotic black holes.

A key issue is the thermodynamical stability of the system. In a holographic construction there is an identification of degrees of freedom on both sides of the duality, so the existence of thermodynamical instabilities should be reflected in the dual geometry along the lines of the Gubser-Mitra conjecture \cite{Gubser:2000ec}. Indeed, the conjecture has been confirmed to hold for backgrounds with finite negative specific heat, $c_V<0$, where there is a dynamically unstable mode that is associated to an imaginary speed of sound in the dual theory \cite{Buchel:2005nt}
$$
v_s^2= \frac{\partial P}{\partial \cE}=\frac{S}{ c_V}\,.
$$
The Hagedorn phase of LSTs is marginally stable, but at weak coupling it can be shown that quantum corrections render it unstable \cite{Kutasov:2000jp, Rangamani:2001ir,Barbon:2007za, LorenteEspin:2007gz}. At strong coupling, black hole solutions that would correspond to LST on compact spaces have been shown to be thermodynamically unstable with negative specific heat \cite{Gubser:2001eg, Bertoldi:2002ks, Buchel:2005nt}. In the flat space limit, the specific heat diverges, $c_V\to -\infty$, so that the speed of sound vanishes and the background is marginally stable \cite{Buchel:2005nt}. Then, according to the Gubser-Mitra conjecture, the geometry should not show any instability at the classical level, although it could be unstable if the effect of quantum corrections were taken into account.

The absence of the tachyon associated to the sound mode was confirmed in \cite{Parnachev:2005hh}. However, the computation of the holographic Green's function \cite{Parnachev:2005hh,Naray
 an:2001dr,DeBoer:2003dd} reveals an analytic structure that seems incompatible with a retarded correlator and maybe introduces other instabilities. If this were the case, the conjecture would fail. Our purpose in this paper is to clarify this issue using a particular setup. We analyze the retarded Green's function in the linear dilaton black hole and compute the spectral function above the mass gap for the operators dual to the dilaton and the metric. We will show that there are no evident instabilities, and explain the origin of the analytic structure found previously. We also find a new set of modes describing energy loss that was not discussed previously.

The outline of the paper is as follows. In section \ref{sec:therm} we will introduce a family of black holes that are the holographic duals to a LST at different volumes and study their thermodynamics. We will concentrate on the infinite volume case and, in section \ref{sec:eoms}, we will give the equations of motion for small metric and dilaton fluctuations and describe some common qualitative features of their effective potentials. In sections \ref{sec:qnms} and \ref{sec:green}, we will discuss the decay of fluctuations in the black hole geometry through quasinormal modes and argue that there are no dynamical instabilities. Then, we will relate the previous results to the scattering matrix and the holographic computation of two-point functions in section \ref{sec:smatrix}. We will present our conclusions in section \ref{sec:conc}. Technical details of the calculations can be found in the appendix.

\section{Thermodynamics of the Hagedorn phase}\label{sec:therm}

LSTs are non-gravitational string theories that describe the dynamics of NS5 branes. They have a density of states that grows exponentially with the energy
$$
\Omega(E) = E^\alpha e^{\beta_H E}\,.
$$
The Hagedorn temperature $T_H=1/\beta_H$ is a limiting temperature of the system. The partition function
$$
Z=\int d E\, \Omega(E)\, e^{-\beta E}
$$
becomes formally divergent beyond this point and a first order phase transition will take place, although the final state is not known for LSTs. The first order phase transition is a characteristic of the canonical ensemble. In the microcanonical ensemble the energy of the system can be increased until the Hagedorn temperature is reached. Then, adding further energy does not increase the temperature and it is invested in exciting internal degrees of freedom instead. To a first approximation, the temperature does not change at all, but, depending on the sign of $\alpha$, there can be a slight slope that gives positive specific heat if $\alpha>0$ or negative specific heat if $\alpha<0$. Thus, the Hagedorn phase could be thermodynamically stable or unstable. It is a highly non-trivial question whether the Hagedorn phase of a given theory would be stable or not.

In order to study the LST at strong coupling we will use a holographic description in terms of an S-dual geometry sourced by a configuration of D5 branes. The solution \cite{Chamseddine:1997nm, Maldacena:2000yy} has been used as the gravitational dual of the theory living on $N_c$ five-branes wrapping an $S^2$. At low energies the theory flows to $SU(N_c)$ $\cN=1$ SYM in flat space \cite{Maldacena:2000yy}. We will consider a deformation of this theory, with $N_f=2 N_c$ five-branes that introduce fundamental degrees of freedom in the dual theory \cite{cnp}. These geometries were constructed in \cite{cnp} and a linear dilaton black hole solution at the Hagedorn temperature of the dual LST was found there as well. As we will discuss later, the presence of the linear dilaton implies that the low energy degrees of freedom of the field theory do not really decouple from the higher LST modes \footnote{At zero temperature there are special cases where there is total decoupling, as was studied in \cite{Caceres:2007mu,HoyosBadajoz:2008fw}.}. Nevertheless, the geometry close to the horizon should still be associated to their holographic description.

In \cite{Bertoldi:2007sf}, a new class of linear dilaton black holes with compact horizon was found with temperature $T$ higher than the
Hagedorn temperature in flat space $T_H = \frac{1}{2\pi \sqrt{N_c \alpha'}}$. The metric in the Einstein frame is
$$
d s^2 = e^{\frac{\phi}{2}}\left[ -f(r) dt^2 + R^2 d\Omega_3^2
+ \frac{N_c\alpha' R^2}{R^2 + N_c} \frac{d r^2}{f(r)} +N_c\alpha' d Y_5^2\right]
\,.
$$
\begin{equation}
f(r)=1-e^{2(r_0-r)}, \ \ \phi = \phi_0 + r\,.
\label{S3bhNNN}\end{equation}
where $r$ is the holographic radial coordinate, $\phi$ is the dilaton and $Y_5$ is the internal space, whose explicit form is unimportant for our analysis. Although quantum corrections will start to be important as $r$ is increased, we are interested in the classical behavior and we will ignore this. In order to remain in a weakly coupled regime, one should perform an S-duality for some value of $r$ and continue along a geometry with decreasing dilaton. This is not very important for our analysis, as will become clearer in section \ref{sec:schrod}.

The horizon is at $r=r_0$ and has a three-sphere $S^3$ component of radius $R={N_h/2}$ proportional to the flux of Ramond-Ramond $3$-form on the sphere
\begin{equation}\label{Rflux}
\int_{S^3} F_{(3)} =  N_h  \,.
\end{equation}
The temperature of these black holes is independent of the radial position of the horizon
\begin{equation}
T(R) = \frac{1}{2 \pi \sqrt{N_c \alpha'}} \, \sqrt{1 + \frac{N_c}{R^2}} = T_H \sqrt{1 + \frac{N_c}{R^2}} > T_H\,,
\label{S3bhT}\end{equation}
We can think of the $T=T_H$ black hole solution with flat $\bR^3$ horizon as the formal limit $R \rightarrow \infty$ of the
solution \eqref{S3bhNNN}
\begin{equation}
\lim_{R \to \infty} T(R) = T_H^+
\end{equation}
This limit is consistent with the vanishing of the Ramond-Ramond $3$-form flux density on the three-sphere,
\begin{equation}
\lim_{R \to \infty} \frac{N_h}{R^3} = \frac{2}{R} \rightarrow 0 \ .
\end{equation}

The entropy of these solutions was computed in \cite{Bertoldi:2007sf}. We compute the energy (\ref{Eren}) and find that
\begin{equation}\label{energyHag}
E=T S = C e^{2 r_0} V(S^3) {T\over T_H}, \ C= N_c^2{e^{2\phi_0}(\alpha')^2 V(Y_5)\over 8\pi G_{10}}\,.
\end{equation}
Here $V(S^3)$ is the volume of the $S^3$ and $V(Y_5)$ the volume of the internal five-dimensional space. Our results agree with the computation
at infinite volume \cite{Cotrone:2007qa} when the $R\to \infty$ limit is taken.

We can interpret each black hole solution as a state of fixed entropy and volume in the {\it microcanonical} ensemble. The temperature is clearly
\begin{equation}
\left({\partial E\over \partial S} \right)_V = T\,.
\end{equation}
If we fix the volume, we can change the energy without changing the temperature by tuning the position of the horizon, so
\begin{equation}
\left({\partial T\over \partial E} \right)_V = 0\,.
\end{equation}
We can use the last result to find the density of states,
\begin{equation}
{1\over T} = {\partial \log \Omega(E) \over \partial E} \Rightarrow \Omega(E) = \Omega_0 e^{E/T}\,.
\end{equation}
Then, all these backgrounds are marginally stable, $\alpha=0$.

\section{Fluctuations in the black hole}\label{sec:eoms}

Small perturbations around the black hole can be studied using the linearized equations of motion in the background.
This allows to study the hydrodynamics of the dual field theory using the methods pioneered in \cite{Lorentzianrecipe} and, more generally,
to evaluate the spectrum of quasinormal modes in the black hole background,  which describe dissipation in the dual theory \cite{qnms,Hoyos:2006gb}.
Hydrodynamic modes in dual LST backgrounds were studied in \cite{Parnachev:2005hh}.

The metric of the flavored $N_f= 2 N_c$ planar black hole is the $R\to \infty$ limit of \eqref{S3bhNNN}
\begin{equation}
d s^2 = e^\frac{\phi}{ 2}\left[ -f(r) dt^2 +d\vec{x}^2+N_c\alpha'\frac{d r^2 }{f(r)} +N_c\alpha' d Y_5^2\right]\,.
\end{equation}
where $t,\vec{x}$ are the coordinates of Minkowsky space.

For simplicity, we will work in units where $\alpha'=1$. We also make the change of variables $r=-(\log u)/2$, so the black hole factor simplifies to $f(u)=1-u$. Notice that $r\to \infty$ corresponds to $u\to 0$.

The metric fluctuations split into three independent channels classified by the $SO(2)$\footnote{For non-zero momentum.} unbroken group of rotations in space: scalar, vector and tensor.
Scalar perturbations mix with the dilaton.

The equations of motion for plane-waves in four-dimensional space are derived in  Appendix \ref{sec:EOMcomputation}. Let us use  dimensionless frequency and momentum in units of the temperature $\wn=\omega/(2\pi T_H)$, $\qn=q/(2\pi T_H)$. From the metric and the dilaton equations, we can build two diffeomorphism-invariant quantities $Z_{sound}$ and $Z_\phi$, obeying
$$
Z_{sound}'' + \frac{\qn^2 u - 2 \wn^2 }{(1-u) ( \qn^2 (u-2) + 2 \wn^2 )} Z_{sound}'
$$
$$
+ \left( \frac{ - 4 u^2 (1-u) \qn^2 +   (u^2 - 3 u + 2) \qn^4 + (3u-4) \qn^2 \wn^2 + 2 \wn^4  }{4 u^2 (1-u)^2 ( \qn^2 (u-2) + 2 \wn^2 )} \right) Z_{sound}
$$
\begin{equation}
- \frac{4 (\qn^4 - 2 \qn^2 \wn^2 )}{3(1-u)( \qn^2 (u-2) + 2 \wn^2 )} Z_\phi' = 0\,,
\label{sound1}\end{equation}
and
\begin{equation}
Z_\phi'' - \frac{1}{1-u} Z_\phi' + \frac{ \wn^2 - (1-u) \qn^2}{4 u^2 (1-u)^2} Z_\phi = 0
\label{dilatonscalarq}\end{equation}
If one sets $Z_\phi$ to zero, equation (\ref{sound1}) becomes the same as the
sound mode equation found in \cite{Parnachev:2005hh} for six-dimensional LST. When the spatial momentum is zero, $\qn=0$,
the two equations decouple and the sound equation reduces to the scalar equation
\begin{equation}
Z_{\phi}'' - \frac{1}{1-u} Z_{\phi}' + \frac{\wn^2 }{4 u^2 (1-u)^2} Z_{\phi} = 0\,.
\label{dilatonscalarq2}\end{equation}
The same equation is also obtained for the tensor component of the metric.

For the shear mode, we find the equation
\begin{equation}
Z_{shear}''(u) + \frac{\wn^2}{(1-u)(\qn^2(1-u)-\wn^2)} Z_{shear}'(u)
- \frac{\qn^2(1-u)-\wn^2}{4(1-u)^2 u^2} Z_{shear}(u) = 0\,.
\label{eq:shear}\end{equation}
All these equations coincide with the equations found in \cite{Parnachev:2005hh} for six-dimensional LST.

%%%%%%%%%%%%%%%%%%%%%%%%%%%%%%5 6d LST

%%%%%%%%%%%%%%%%%%%%%%%%%%%%%%%%%
\subsection{Schr\"odinger potentials for classical modes}\label{sec:schrod}

In order to understand better the asymptotic conditions on the solutions to the equations of motion, it is convenient to write them in a Schr\"odinger form
using a Regge-Wheeler tortoise coordinate. The resulting  ``effective potentials" can also be used to study the 1-loop corrections
to the black hole free energy due to supergravity modes. This was done in \cite{Barbon:2007za} for the system of flat $D5$-branes at weak coupling. In that case,
1-loop corrections lift the Hagedorn degeneracy of the black hole.

Let us take $Z_\phi(u)=\sqrt{u} \psi(u)$ and change variables to the Regge-Wheeler tortoise coordinate $v=\log(u)-\log(1-u)$.
In this new coordinate, spatial infinity corresponds to $v\to - \infty$, while the black hole horizon is at $v\to\infty$.
Equation (\ref{sound1}) has now the form of a Schr\"odinger equation $\left(-\partial_v^2 +V(v)\right)\psi = E \psi$, where the energy is
\begin{equation}\label{eq:enepot}
E=\frac{\wn^2}{ 4}
\end{equation}
and the potential is
\begin{equation}
    V(v)= \frac{1-u^2(v) + \qn^2(\, 1- u(v) \,)}{ 4}
    \label{eq:schpot2}
\end{equation}
The potential is positive and monotonic, and it is equal to $(1 + \qn^2)/4 $ at infinity and vanishes
at the horizon. Solutions can be expanded in ingoing and outgoing plane waves at the horizon. At infinity there are two possible situations.
For $E<(1+ \qn^2)/4$ or $\wn^2- \qn^2 < 1$,  one solution blows up and the other is exponentially decreasing. On the other hand,
for $E > (1+ \qn^2)/4$ or $\wn^2- \qn^2 > 1$, the two solutions correspond to outgoing or ingoing plane waves.

We can repeat the same analysis for the sound and the shear mode, obtaining similar results
$$
\begin{array}{lcr}
Z_{\rm sound} = \sqrt{u} ( \qn^2(u-2) + 2 \wn^2 ) \psi & , & Z_{\rm shear} = \sqrt{u ( \wn^2 - \qn^2(1-u) )} \ \psi \\
\end{array}
$$
\begin{eqnarray*}
  V_{\rm sound}(v) & = & \frac{1-u^2(v) + \qn^2 (\, 1 - u(v) \,) }{ 4}
+ \frac{2 \qn^2  (2\wn^2-\qn^2) (1-u(v)) u^2(v)}{\left( \qn^2(u(v)-2)+2\wn^2 \right)^2}\\
V_{\rm shear}(v) &  = & \frac{1-u^2(v) + \qn^2 (\, 1- u(v) \,)}{4}
+ \frac{\qn^2 (1-u(v)) u^2(v) ( \qn^2 (1-u(v)) + 2 \wn^2 )}
{4 \left( q^2(1 - u(v))-\wn^2 \right)^2}
\end{eqnarray*}

The potentials have similar asymptotics. The behavior close to the horizon is typical of black holes, while the asymptotics at infinity are due to the non-trivial dilaton profile.
The height of the potential is given by the slope of the linear dilaton background and it is independent of the sign, so that the same potential will be found for
the S-dual geometry as we had anticipated.

We have computed the equations and the potentials explicitly for the $N_f=2 N_c$ background, but other black hole backgrounds asymptotically conformally equivalent to flat space
and with a linear dilaton will have the same asymptotic behavior of the effective potential, flat and vanishing at the horizon and constant at large values of the radial coordinate, e.g. \cite{Parnachev:2005hh,Narayan:2001dr,DeBoer:2003dd}.

Therefore, we expect that the following analysis will produce qualitatively similar results in more general cases.

\section{Quasinormal modes and speed of sound}\label{sec:qnms}

Localized fluctuations in a black hole background tend to spread and disappear as time passes. The energy that initially is concentrated in some finite region falls into the black hole
or escapes to infinity, so that the local energy density vanishes. This process is described by the quasinormal modes (QNMs). In order to define QNMs,
suitable boundary conditions have to be imposed at the horizon and at spatial infinity. The condition at the horizon is that the mode should be infalling.
In terms of the effective potentials computed in section \ref{sec:schrod}, this means that we pick the plane wave moving towards the horizon $v\to \infty $
$$
\sim e^{-i\wn t+i \wn v/2}.
$$
At spatial infinity $v\to -\infty$, we have two possible situations, depending on the four-dimensional mass shell $\wn^2-\qn^2$. If $\wn^2-\qn^2 < 1$, then the
 modes are below the potential barrier, so we must pick the vanishing solution at infinity (Dirichlet boundary condition). However, if $\wn^2-\qn^2 > 1$,
 then the modes are  above the barrier and the right boundary condition is to pick the outgoing plane wave. The asymptotic form of the mode is then
$$
\begin{array}{ll}
\wn^2-\qn^2 < 1 & \sim e^{-i\wn t + \sqrt{1+\qn^2-\wn^2} v/2} \\
\wn^2-\qn^2 > 1 & \sim e^{-i\wn t-i  \sqrt{\wn^2-\qn^2-1} v/2}
\end{array}
$$
For ${\rm Im}\, \wn < 0 $, the QNMs describe fluctuations that decrease exponentially in time, hence their association with dissipation.
On the other hand, modes with ${\rm Im}\, \wn > 0 $ would be in principle interpreted as instabilities of the background.
The quasinormal spectrum of the dilaton and shear modes can be obtained analytically, because their equation of motion can be reduced to a hypergeometric equation. Although this is not the case for the sound mode, we are able to show that the speed of sound vanishes, $v_s=0$ (\ref{eq:vs}).
This is in agreement with our expectations from the thermodynamical analysis: a vanishing speed of sound signals a marginal (in)stability.

\subsection{Solutions of the hypergeometric equation}\label{sec:hyperg}

In this section we will establish the notation that will be used in the rest of the paper. Consider the hypergeometric equation
\begin{equation}
    u(1-u) y''+(c-(a+b+1)u) y'- a b  y =0\,,
    \label{eq:hypergeom}
\end{equation}
The local solutions close to the singular points are given in terms of hypergeometric functions $_2F_1(a,b,c;x)=1+ \frac{a b}{c} x + O(x^2)$, $x\to 0$. At spatial infinity ($u\to 0^+$),
\begin{eqnarray}
\notag  y(u) & = & a_0 y^{(0)}_0(u) + b_0 y^{(1)}_0(u) \, , \\
\notag    y^{(0)}_0(u)& = &  _2F_1(a,b,c;u)\,,\\
 y^{(1)}_0(u) & = & u^{1-c}\, _2F_1(a+1-c,b+1-c,2-c;u)\,.
    \label{eq:solinfinity}
\end{eqnarray}
At the horizon ($u\to 1^-$), the local solutions are
\begin{eqnarray}
\notag  y(u) & = & a_1 y^{(0)}_1(u) + b_1 y^{(1)}_1(u) \,.\\
\notag    y^{(0)}_1(u)& = &  _2F_1(a,b,1-c+a+b;1-u)\,,\\
 y^{(1)}_1(u) & = &  (1-u)^{c-a-b} \, _2F_1(c-a,c-b,1+c-a-b;1-u)\,.
    \label{eq:solhorizon}
\end{eqnarray}
Both sets of solutions are related by the connection coefficients
\begin{equation}
    \left(\begin{array}{c} y^{(0)}_1 \\ y^{(1)}_1 \end{array}\right) = \left( \begin{array}{cc} C_{00} & C_{01} \\ C_{10} & C_{11} \end{array}\right)
    \left(\begin{array}{c} y^{(0)}_0 \\ y^{(1)}_0  \end{array}\right)\,,
    \label{eq:connect2}
\end{equation}
whose explicit analytic form is
\begin{eqnarray}
C_{00} & = & \frac{\Gamma(1-c+a+b) \Gamma(1-c)}{\Gamma(1-c+b) \Gamma(1-c+a)} \nonumber \\
C_{01} & = & \frac{\Gamma(1-c+a+b) \Gamma(c-1)}{\Gamma(a) \Gamma(b)} \nonumber \\
C_{10} & = & \frac{\Gamma(1+c-a-b) \Gamma(1-c)}{\Gamma(1-a) \Gamma(1-b)} \nonumber \\
C_{11} & = & \frac{\Gamma(1+c-a-b) \Gamma(c-1)}{\Gamma(c-b) \Gamma(c-a)}\,.
\label{connectionHyper2}\end{eqnarray}
When the values of $a$, $b$ and $c$ are such that a connection coefficient vanishes, then one of the hypergeometric solutions becomes a polynomial. On the other hand, for the values where the connection coefficient has a pole, the hypergeometric series diverges.

\subsection{Dilaton and tensor mode}

The equation of the dilaton with nonzero momentum, \eqref{dilatonscalarq}, reduces to a hypergeometric equation via
\begin{equation}\label{eq:hypergphi}
Z_\phi[u]=u^{c/2}(1-u)^{-i\wn/2} y(u)
\end{equation}
where the parameters of the hypergeometric equation are
\begin{equation}\label{eq:cdilaton}
c=1+\sqrt{1+\qn^2-\wn^2}
\end{equation}
and
\begin{equation}\label{eq:abdilaton}
a = b = \frac{c}{2} -i \frac{\wn}{2} = \frac{1  + \sqrt{1+\qn^2-\wn^2}- i \wn}{2}
\end{equation}
With this choice for the dilaton perturbation \eqref{eq:hypergphi}, the infalling boundary condition corresponds to $y(u)\sim 1$ close to the horizon, $u\to 1^-$, $v \to + \infty$, that is the $y_1^{(0)}$ solution
$$
Z_\phi \sim (1-u)^{-i\wn/2} y_1^{(0)}(u) \sim e^{i \wn v/2}
$$
The Dirichlet condition at spatial infinity, $u \to 0^+$, $v \to -\infty$ corresponds to $y(u)\sim 1$, ($y_0^{(0)}$)
$$
Z_\phi \sim u^{c/2} y_0^{(0)} \sim \sqrt{u} \ \psi^+_<(\wn,\qn,v)
$$
where
$$
\psi^+_<(\wn,\qn,v) \rightarrow e^{\sqrt{1+\qn^2-\wn^2}\, v/2}
$$
is the normalizable solution, for values $\Im\wn =0$, $\wn^2 - \qn^2 < 1$, corresponding to the principal branch of the square root $\Re\sqrt{1+\qn^2-\wn^2} > 0$. One can choose to set the branch cut for $\sqrt{1+\qn^2-\wn^2}$ at $\Im\wn =0$, $\wn^2 > \qn^2+1$, with
$\Re\sqrt{1+\qn^2-\wn^2} > 0$ on the sheet that includes the physical frequency domain and $\Re\sqrt{1+\qn^2-\wn^2} < 0$ on the second sheet. This choice implies that next to the branch cuts,
\begin{equation}\label{eq:branches}
\sign(\Im\sqrt{1+\qn^2-\wn^2}) = \mp\sign (\Re\wn), \ \ \Im\wn\to 0^\pm\,.
\end{equation}
on the first sheet and the opposite signs on the second sheet. The sign of the imaginary part of the square root for each sheet is represented in figure \ref{branches}.

\FIGURE[ht!]{
\includegraphics[width=12cm]{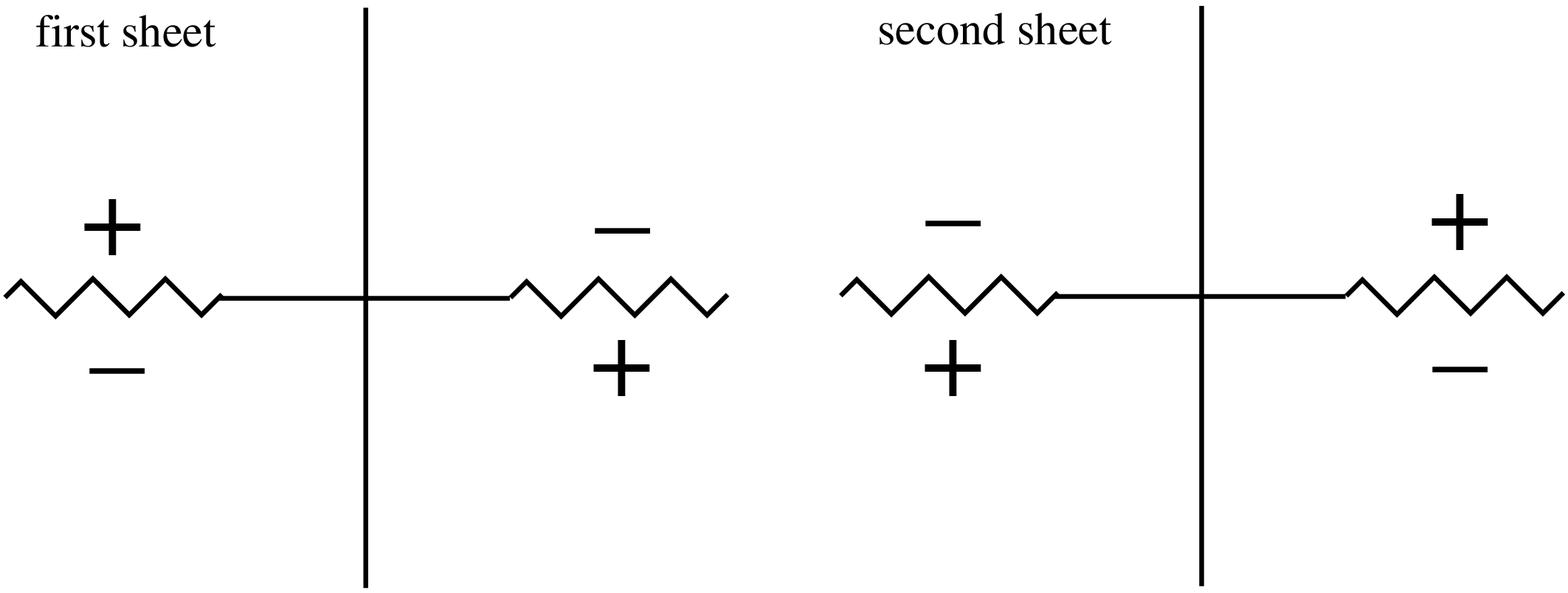}
\caption{\label{branches} Branch cuts of $\sqrt{1+\qn^2-\wn^2}$ on the complex $\wn$ plane. Branch points are at $\wn=\pm \sqrt{1+\qn^2}$, the first sheet corresponds to  $\Re\sqrt{1+\qn^2-\wn^2} > 0$ and the second sheet to  $\Re\sqrt{1+\qn^2-\wn^2} < 0$. The signs above and below the branch cuts correspond to the sign of $\Im\sqrt{1+\qn^2-\wn^2}$.}
}

The solution is initially defined on a real line above the branch cuts $\Im\wn\to 0^+$. From \eqref{eq:branches} this implies that $\psi^+_<$ also corresponds to the outgoing solution, for $\wn^2>\qn^2+1$. Indeed, when $\Re\wn > \sqrt{1+\qn^2}$,
\begin{equation}
e^{-i\wn t} \psi^+_< \sim e^{-i\Re\wn\, t + i \Im\sqrt{1+\qn^2-\wn^2} \, v/2}
\end{equation}
so the solution corresponds to a plane wave moving towards infinity $v\to-\infty$ and  it is straightforward to see that the same holds for $\Re\wn <0$.
%It is possible to do a different choice of branch cut, and although the details of the computation are different, the physical result is the same.

Therefore, we find that quasinormal modes correspond to solutions where $y_1^{(0)}$ is
proportional to $y_0^{(0)}$.
The condition is that the connection coefficient $C_{01}=0$. Using \eqref{connectionHyper2} and \eqref{eq:abdilaton}
\begin{equation}\label{eq:noqnms}
a=\frac{1+\sqrt{1+\qn^2-\wn^2}-i\wn }{ 2} = -n, \ \ n=0,1,2,\dots
\end{equation}
There are no solutions to this equation on the physical sheet. This agrees with previous results found in related Little String backgrounds \cite{Parnachev:2005hh}.

It is possible to do an analytic extension of the solution to the second sheet. This is equivalent to a different choice of solution at spatial infinity
$$
Z_\phi \sim u^{c/2} y_0^{(1)} \sim \sqrt{u} \ \psi^-_<(\wn,\qn,v)
$$
In this case quasinormal modes correspond to solutions where $y_1^{(0)}$ is
proportional to $y_0^{(1)}$. The condition is that the connection coefficient $C_{00}$ vanishes
\begin{equation}\label{eq:qnmdirichlet}
1-c+a=\frac{1-\sqrt{1+\qn^2-\wn^2}-i\wn }{ 2} = -n, \ \ n=0,1,2,\dots
\end{equation}
This condition is defined now on the first sheet of the square root, notice that it can be obtained from \eqref{eq:noqnms} by flipping the sign of the square root. In this case there is an infinite discrete set of solutions
\begin{equation}\label{eq:qnmspectrum}
\wn=-2 i \frac{n(n+1)}{ (2n+1)}+i \frac{\qn^2}{ (2(2n+1))}, \ \ n=0,1,2,\dots\,.
\end{equation}
For $\qn=0$ the $n=0$ mode is actually not in the spectrum ($C_{00} \neq 0$) and the full set of quasinormal modes have ${\rm Im}\, \wn < 0 $. When $\qn\neq 0$, the $n=0$ mode appears above the real axis $\wn \sim +i \qn^2/2$ and as we increase the momentum, more modes move to the upper half plane.

\subsection{Shear mode}

We have given an equation for the shear mode \eqref{eq:shear} in terms of the gauge-invariant quantity $Z_{shear}$.
For the analysis of the quasinormal modes we will work instead with non-invariant quantities $H_{tx}'$ and $H_{zx}'$.
The location of the quasinormal modes is itself gauge-invariant, so it is irrelevant which variable we are using.
As a consistency check, we will see that $H_{tx}'$ and $H_{zx}'$ have the same quasinormal spectrum. The equations for the shear variables are
$$
\wn H_{tx}'(u) + \qn(1-u) H_{zx}'(u) = 0 \ ,
$$
$$
H_{zx}''(u)
- \frac{H_{zx}'(u)}{(1-u)}
+ \frac{\qn \,\wn H_{tx}(u) +\wn^2 H_{zx}(u)}{4(1-u)^2 u^2} = 0 \ ,
$$
\begin{equation}
H_{tx}''(u)
- \frac{\qn^2 H_{tx}(u) + \qn\, \wn H_{zx}(u)}{4(1-u) u^2} = 0 \ ,
\end{equation}
We can solve for $H_{zx}'$ from the first equation and for $H_{zx}$ from the third. Deriving the third equation with respect to $u$ and plugging back the expressions for $H_{zx}$ and $H_{zx}'$,
we find a hypergeometric equation for $H_{tx}'$. Similarly, we can solve for $H_{tx}'$ in the first equation and for $H_{tx}$ in the second.
Then, deriving the second equation with respect to $u$ and substituting the expressions of $H_{tx}$ and $H_{tx}'$, we find the same hypergeometric equation for $H_{zx}'$.
This shows that the quasinormal spectrum is the same for both.

The parameters of the hypergeometric equation are
\begin{equation}\label{eq:cshear}
c=1+\sqrt{1+\qn^2-\wn^2}\,,
\end{equation}
and
\begin{eqnarray}\label{eq:abshear}
\notag a & = &  \frac{c}{2}-1 -i \frac{\wn}{2} = \frac{-1  + \sqrt{1+\qn^2-\wn^2}- i \wn}{2} \\
b & = &  \frac{c}{2}+1 -i \frac{\wn}{2} = \frac{3  + \sqrt{1+\qn^2-\wn^2}- i \wn}{2}\,.
\end{eqnarray}
The analysis follows in the same way as for the dilaton.
The quasinormal modes are determined by the conditions
\begin{eqnarray}
\notag a & = & -n, \ \ n=0,1,2,\dots \\
b & =& -n, \ \ n=0,1,2,\dots
\end{eqnarray}
The solution of $a=0$ is a mode with $\wn=-i \qn^2/2$, which is precisely the dispersion relation of a hydrodynamic mode in the holographic dual. Apart from this mode, the above system of equations yields the same quasinormal spectrum that we found for the dilaton \eqref{eq:qnmspectrum}. Therefore,
modes in the upper half plane appear for finite values of the momentum, but these modes
with $\Im \wn > 0$ are all on the second sheet.

%%%%%%%%%%%%%%%%%%%%%%%
\subsection{The speed of sound}\label{sec:vs}

Let us set the dilaton to zero. The sound equation is a more complicated equation with four regular singularities at $u=0,1,2-2\wn^2/\qn^2, \infty$.
The change $Z=u^{\rho/2} (1-u)^{\sigma/2} y(u)$, with $\rho=1+\sqrt{1+\qn^2-\wn^2}$ and $\sigma =-i\wn$ gives the Heun equation
\begin{equation}
y''+\left(\frac{\gamma}{ u}+\frac{\delta}{ u-1}+\frac{\epsilon }{ u-2+2\wn^2/\qn^2}\right) y' +\frac{\alpha \beta u -Q }{ u (u-1)(u-2+2 \wn^2/\qn^2)}y =0
\end{equation}
where
\begin{eqnarray}
\notag \gamma & = & \rho = 1+\sqrt{1+\qn^2-\wn^2} \\
\notag \delta & = & \sigma+1 = 1-i\wn \\
\notag \epsilon & = & -2 \\
\notag \alpha & = & \beta\ =\ \frac{1}{ 2} (\gamma+\delta+\epsilon-1)\ =\ \frac{1}{ 2}(-1-i \wn +\sqrt{1+\qn^2-\wn^2 } ) \\
\notag Q & = & \frac{\qn^2}{ 2} - {\wn\over 2} (3\wn +2 i (1+\sqrt{1+\qn^2-\wn^2})) \\ \notag \ & \ & +{\wn^2\over \qn^2}(-1+ \wn^2-\sqrt{1+\qn^2-\wn^2}
+i\wn (1+\sqrt{1+\qn^2-\wn^2}))\\
 \ & \ & \,
\end{eqnarray}
There is an explicit constant solution $y(u)=1$ corresponding to the sound mode. Such solution appears when $\alpha=\beta=Q=0$, that in this case is satisfied for the dispersion relation
\begin{equation}\label{eq:vs}
\wn=-i{\qn^2 \over 2}\,.
\end{equation}
Therefore, the speed of sound is zero, $v_s=0$, in the $N_f=2 N_c$ holographic dual theory.

\section{Retarded two-point function in the black hole}\label{sec:green}

We have found a set of QNMs with ${\rm Im}\, \wn > 0 $  for $\qn>0$. In order to clarify the issue of the QNMs with positive imaginary part, we will follow a similar analysis to \cite{QNMs}. We consider the Fourier transform of solutions of the Schr\"odinger equations introduced in section 3 for times $t>0$
\begin{equation}\label{eq:laplacetransf}
\psi(\wn,\qn,v)  =  \int_0^\infty e^{i\wn t} \psi(t,\qn,v) dt, \ \  \Im \wn > 0 \,.
\end{equation}
As in \cite{QNMs} the solution $\psi(\wn,\qn,v)$ can be thought of as the evolution by an appropriate Green's function of a perturbation that starts at $t=0$. The Green's function $G(\wn,\qn;v,v')$  is determined by the choice of two linearly independent solutions of
the Schr\"odinger equation, $\psi_<$ and $\psi_>$, that satisfy the appropriate boundary conditions
at $v = - \infty$ and $v = + \infty$ respectively. The explicit form is
\begin{equation}\label{eq:greenf}
G(\wn,\qn;v,v') = \frac{\psi_<(\wn,\qn,v_<) \psi_>(\wn,\qn,v_>)}{W(\wn,\qn)}
\end{equation}
where $v_< = {\rm min}(v,v')$, $v_> = {\rm max}(v,v')$ and $W(\wn,\qn)$ is the Wronskian of the two solutions,
\begin{equation}\label{wrons1}
i W(\wn,\qn) = \psi_<(\wn,\qn,v) \frac{\partial}{\partial v} \psi_>(\wn,\qn,v) - \frac{\partial}{\partial v} \psi_<(\wn,\qn,v) \psi_>(\wn,\qn,v),
\end{equation}

As we saw in section \ref{sec:schrod}, the potential vanishes at $v = + \infty$ and asymptotes to
$(1 + \qn^2)/4$ at $v = - \infty$. Physical solutions $\psi(\wn,\qn)$ should be at least delta-normalizable in the bulk. This imposes the following boundary condition as $v \to + \infty$:
$$
\psi^+_>(\wn,\qn,v) \sim e^{i\wn v/2}\,.
$$
Notice that this is the prescription to compute a retarded correlator, since the black hole does not radiate classically. Conversely, as $v \to - \infty$, we find that solutions can be a linear combination of
$$
\psi^\pm_<(\wn,\qn,v) \rightarrow e^{\pm \sqrt{1+\qn^2-\wn^2}\, v/2}
$$
For values $\Im\wn =0^+$, $\wn^2 < \qn^2+1$, corresponding to $\Re\sqrt{1+\qn^2- \wn^2} > 0$, only $\psi^+_<$ is normalizable. For values of the frequency $\wn^2 > \qn^2+1$, $\psi^+_<$ also corresponds to an outgoing plane wave, as was explained in section \ref{sec:qnms}.

The retarded bulk correlator \eqref{eq:greenf} for the dilaton and the scalar modes has the following form
\begin{equation}\label{bulkcorrelator}
G(u,u')=i\frac{ (u_< u_>)^{(c-1)/2} [ (1-u_<) (1 - u_>) ]^{- i \wn/2}}{(1-c) C_{01}}\,{}_2F_1 (a,b,c;u_<) {}_2 F_1(a,b,1-c+a+b;1-u_>) \ ,
\end{equation}
where we used the fact that $\psi^+_<  = u^{(c-1)/2} (1 - u)^{- i \wn/2}  y_0^{(0)}$
and $\psi^+_> = u^{(c-1)/2} (1 - u)^{- i \wn/2} y_1^{(0)}$.

Since $\psi^+_<$ and $\psi^+_>$ are solution of a Schr\"odinger equation, their Wronskian \eqref{wrons1}
is constant. This constant can be evaluated by considering the limit $u \to 0$.
Using
$$
\frac{\partial}{\partial v} = (1-u) u \frac{\partial}{\partial u}
$$
and
$$
y_1^{(0)} = C_{00} y_0^{(0)} +C_{01} y_0^{(1)}
=   C_{00}\  {}_2F_1(a,b,c;u) + C_{01}\, u^{1-c}\, _2F_1(a+1-c,b+1-c,2-c;u)\,
$$
we find
\begin{equation}\label{wronskian}
i W
= u^{c} (1 - u)^{1- i \wn}  C_{01} \left[
y_0^{(0)} \partial_u  y_0^{(1)}
- y_0^{(1)} \partial_u  y_0^{(0)}
\right]
= (1 - c) C_{01}
= - \frac{\Gamma(1-c+a+b) \Gamma(c)}{\Gamma(a) \Gamma(b)}\ ,
\end{equation}
which leads to \eqref{bulkcorrelator}.

The Wronskian vanishes when the connection coefficient between the two solutions, $\psi^+_<$ and
$\psi^+_>$, has a zero. The inverse of the Wronskian (for the scalar mode) is proportional to
\begin{equation}\label{scalarW}
W(\wn,\qn)^{-1}  \sim C_{01}^{-1} = {\Gamma\left( \frac{1 + \sqrt{1 + \qn^2 -\wn^2} -i\wn}{2} \right)^2\over \Gamma(\sqrt{1+\qn^2-\wn^2}) \Gamma(1-i\wn)}\,.
\end{equation}
An important subtlety in the Green's function are the singularities of the solutions. The solutions are proportional to hypergeometric functions whose series coefficients diverge for some complex values of $\wn$, implying that there is no solution. This happens when $c<0$ or $1+a+b-c<0$ in the first sheet or $2-c<0$ or $1-a-b+c<0$ in the second sheet. The bad frequencies are
$$
\wn =-i n,\ \ \wn^2 =1-n^2 +\qn^2, \ \ n=1,2,\dots
$$
This is taken into account by the singularities of the Wronskian, that cancel against any such spurious singularities. Notice that those solutions correspond to the poles in the overall factor, as can be seen in \eqref{wronskian}
\begin{equation}
W \sim \Gamma(1-c+a+b) \Gamma(c)
\end{equation}
The singular solutions are for instance at the origin of singularities on the real axis that were found in the holographic computation of the dual theory Green's functions \cite{Parnachev:2005hh,Narayan:2001dr,DeBoer:2003dd}.

Then, the only genuine singularities come from the zeroes of the Wronskian, and one can show that with the above choice of solutions $\psi^+_<$ and
$\psi^+_>$, the corresponding Green's function has no singularities when it is analytically continued on the $\Im\wn >0$ half plane \eqref{eq:noqnms}. Therefore,
this Green's function can be identified with the retarded correlator.

The full analytic extension of the Green's function includes the $\Im \wn <0$ half plane and the second Riemann sheet that can be reached through the square root branch cuts $\Im\wn =0$, $\wn^2 > \qn^2+1$. This amounts to flipping the sign of the square root in the Wronskian (\ref{scalarW}), which is equivalent to changing the $\psi^+_<$ solution to
$\psi^-_<$.

Singularities on the $\Im \wn <0$ half plane on either sheet correspond to solutions of the Schr\"odinger equation that diverge as $v\to\infty$ and that can be associated to absorption by the black hole at late times, as for the usual QNMs studied in AdS spaces. In this case, all the poles are on the second sheet
$$
\wn_n=-2 i \frac{n(n+1)}{ (2n+1)}+i \frac{\qn^2}{ (2(2n+1))}, \ \ n=1,2,\dots\,.
$$
Singularities on the upper half plane of the second sheet $\Im \wn >0$, correspond to modes that diverge as $v\to -\infty$, so they also describe the loss of energy in the bulk but as it escapes towards spatial infinity. This is clearly different from QNMs in AdS spaces, where the geometry acts effectively as a box and the energy loss is produced only by absorption by the black hole.

In summary, it is possible to define a retarded Green's function, analytic in the
upper half frequency plane of the physical sheet. The QNMs with $\Im\wn >0$ are all in the second sheet, so they are related to processes with no black hole absorption and where all the energy escapes to infinity.

For the shear mode a similar analysis leads to
\begin{equation}
W(\wn,\qn)^{-1} \sim {\Gamma\left( \frac{-1  + \sqrt{1+\qn^2-\wn^2}- i \wn}{2} \right)\Gamma\left( \frac{3  + \sqrt{1+\qn^2-\wn^2}- i \wn}{2} \right) \over \Gamma(\sqrt{1+\qn^2-\wn^2}) \Gamma(1-i\wn)}\,.
\end{equation}
The spectrum of QNMs is the same on the second sheet. On the first sheet, there is a single pole associated to the shear diffusion in the dual theory
$$
\wn=-i{\qn^2\over 2}\,.
$$

\section{Scattered waves and holographic spectral function}\label{sec:smatrix}

We now understand the physical meaning of the unusual (as compared to AdS) QNMs from the perspective of the linear dilaton geometry. We now want to interpret them in terms of the holographic theory. The simplest approach is to use the relation between the greybody factors for the fields in the black hole geometry and the spectral function of the dual theory \cite{Gubser:1997cm,Gubser:1997se}.

When the black hole is at equilibrium with a thermal gas, the emitted radiation is the same as the absorbed radiation, so the greybody factor is simply the transmission coefficient. Then, we have to solve the scattering problem of radiation coming from spatial infinity. The scattering problem in the context of zero temperature linear dilaton backgrounds was studied previously in \cite{Minwalla:1999xi}, where a relation between the scattering matrix and holographic Green's functions was proposed. In this case, there are real pole singularities that are associated to normalizable states and that should be extracted from the scattering amplitude in order to compute the Green's function, following a LSZ reduction as the one proposed in \cite{Aharony:2004xn}. Normalizable states were localized at the bottom of the linear dilaton geometry, that in our case has been swallowed by the black hole. These states have disappeared from the spectrum and the reduction is not necessary in our
  case.

For $\wn>\sqrt{\qn^2+1}$, radiation enters the black hole from spatial infinity and is partially reflected and partially transmitted
\begin{equation}
\begin{array}{lcl}
v\to \infty & , & \ \ e^{i \wn v/2} \\
v\to -\infty & , & \ \ A^+_{\rm in}  e^{i  \sqrt{\wn^2-\qn^2-1} v/2} + A^+_{\rm out} e^{-i  \sqrt{\wn^2-\qn^2-1} v/2}
\end{array}
\end{equation}
where $A^+_{\rm in}=C_{00}$ and $A^+_{\rm out}=C_{01}$ are  the connection coefficients of the hypergeometric equation given in the section \ref{sec:qnms}.

For negative frequencies $\wn<-\sqrt{\qn^2+1}$, the scattered waves should describe the time reversal of the absorption process
\begin{equation}
\begin{array}{lcl}
v\to \infty & , & \ \ e^{-i \wn v/2} \\
v\to -\infty & , & \ \ A^-_{\rm out} e^{i  \sqrt{\wn^2-\qn^2-1} v/2} + A^-_{\rm in}  e^{-i  \sqrt{\wn^2-\qn^2-1} v/2}\end{array}
\end{equation}
where $A^-_{\rm in}=C_{11}$ and $A^-_{\rm out}=C_{10}$.

Comparing the 'probability current' $J=-i(\psi^*\partial_v \psi -\,{\rm c.c.}\,)$ at the boundary and at the horizon,
we can define the transmission and reflection coefficients. From the expression of the connection coefficients (\ref{connectionHyper2}),
and using the relations
$$
\Gamma(1+i x) \Gamma(1-i x) = {\pi x \over \sinh (\pi x)}, \ \ \Gamma\left({1\over 2}+i x\right) \Gamma\left({1\over 2}-i x\right) = {\pi  \over \cosh (\pi x)}, \ \ \Gamma(1+x)=x \Gamma(x)\,
$$
it turns out that the coefficients for scalar and shear perturbations are the same
\begin{equation}\label{eq:refl}
\cR^\pm = {|A^\pm_{\rm out}|^2\over |A^\pm_{\rm in}|^2} = {\cosh^2\left[{\pi\over 2}(\sqrt{\wn^2-\qn^2-1}\mp\wn)\right]\over \cosh^2\left[{\pi\over 2}(\sqrt{\wn^2-\qn^2-1}\pm\wn)\right]}
\end{equation}
\begin{equation}\label{transmission}
\cT^\pm =  {\pm \wn\over \sqrt{\wn^2-\qn^2-1} |A^\pm_{\rm in}|^2}  = 1- \cR^\pm= {\pm \sinh(\pi\wn) \sinh(\pi\sqrt{\wn^2-\qn^2-1} ) \over \cosh^2\left[{\pi\over 2}(\sqrt{\wn^2-\qn^2-1}\pm \wn)\right]}
\end{equation}
Perturbations close to the mass-shell of the Hagedorn temperature $\wn^2-\qn^2 \simeq 1$ suffer a very small absorption by the black hole $\cR^\pm \simeq 1$, $\cT^\pm\simeq 0$.
Higher mass perturbations on the other hand are easily absorbed, since $\cR^\pm\to 0$ and $\cT^\pm\to 1$, with a large mass limit ($\wn\to\pm \infty$)
\begin{equation}
\cR^\pm \simeq {1\over \cosh^2(\pi \wn)}\to 0, \ \ \cT^\pm \simeq \tanh^2(\pi\wn)\to 1\,.
\end{equation}

 Using the expression of the transmission coefficients $\cT^\pm(\wn)$, we can write down
the spectral function for the dual theory.
Assuming time reversal invariance, the spectral function should be an odd function of the frequency,
$\rho(\wn)=-\rho(-\wn)$, and it should be positive for positive frequency. Then,
\begin{equation}\label{eq:spectral}
\rho_{\rm LST}(\wn) \propto \cT^+(\wn) \Theta(\wn-\sqrt{1+\qn^2}) - \cT^-(\wn)\Theta(-\wn-\sqrt{1+\qn^2})\, ,
\end{equation}
where $\Theta(x)$ is the unit step function. Notice that the spectral function vanishes below the mass gap and that there is no discontinuity since $\cT^\pm(\pm \sqrt{1+\qn^2})=0$. This follows from our analysis, and otherwise the spectral function would be ill-defined, since the factors $\cT^\pm$ are oscillatory below the mass gap. Notice also that the zeroes coincide with the values where there are no solutions in the geometry. Finally, although the explicit form of \eqref{transmission} has the information about the values of the QNMs that describe energy loss towards spatial infinity, their interpretation as resonances is unclear, since the spectral function does not admit a simple analytic extension outside the real line.
%, which may be a consequence of the non-locality of the LST.

If the gap is sent to zero, the spectral function becomes smooth
$$
\rho_{LST}(\wn) \sim \tanh^2(\wn)\,,
$$
while for a very large gap $M$ or a large $\wn$, it is almost like a step function
$$
\rho_{LST}(\wn) \sim  \Theta(\wn-M) - \Theta(-\wn-M)\,,
$$
recovering the zero temperature behavior. So, in the Hagedorn phase there is a general suppression, especially of the states close to the zero momentum mass gap.

The QNMs that contribute to this spectral function correspond to those in the second sheet of the retarded Green's function in the black hole. Only the hydrodynamic modes are not captured by the little string theory degrees of freedom. They can be seen in computations of the retarded Green's function of the field theory using the boundary action for modes vanishing at spatial infinity, as in \cite{Parnachev:2005hh,Narayan:2001dr,DeBoer:2003dd}. However, the Green's function computed in this way presents spurious singularities on the real and imaginary axis. This suggests that the usual holographic prescription for the computation of correlators should be modified in this case.

\section{Discussion}\label{sec:conc}

The linear dilaton black hole shows a quite peculiar analytic structure in its correlators due to the mass gap. In other holographic setups like AdS/CFT, the retarded Green's functions only have singularities in the lower half frequency plane. Exceptions to this rule when a quasinormal frequency crosses the real axis to the upper half of the complex plane translate into dynamical instabilities. This is the case of the instabilities appearing for near-critical embeddings of flavor D7 branes in $AdS_5$  as the quark mass is varied \cite{Hoyos:2006gb,Mateos:2007vn,Myers:2007we}, the unstable sound mode of backgrounds above the Hagedorn temperature \cite{Buchel:2005nt} or instabilities of the Gregory-Laflamme kind
\cite{Gregory:1993vy}.

The reason is that while a quasinormal mode usually has a divergent behavior at spatial infinity, when it crosses the real axis it becomes a bounded state in the geometry. It has finite energy and thus belongs to the physical spectrum, but the energy is negative, which means that  it corresponds to an instability. In the linear dilaton geometry, the modes that cross the real axis always blow up at spatial infinity, so they never belong to the physical spectrum and do not produce any instability. A technical way to see this is that they are on a 'second sheet' of the retarded Green's function. Other apparent instabilities, including poles on the real axis, are actually not present. They correspond to special values of the frequency where there are no well-defined solutions to the equations of motion.

This solves the apparent inconsistency of the Gubser-Mitra conjecture for holographic constructions with the results found in \cite{Parnachev:2005hh,Narayan:2001dr,DeBoer:2003dd}. The marginal stability of LST at the Hagedorn temperature is confirmed by the dynamical stability of the dual linear dilaton geometry. Of course, quantum corrections can drastically change this result, but this is beyond the realm of classical supergravity, which was our main interest. The holographic computation of the spectral function above the mass gap confirms this picture, and no instability appears.

The discrepancy between the bulk analysis and the holographic computation of Green's function using the boundary action suggests that the latter needs to be modified. A possible way would be to use the bulk Green's function to construct a properly defined bulk-to-boundary propagator. The holographic Green's function could then be found from the convolution of two such propagators in the bulk or by finding the boundary-to-boundary limit. In the presence of the black hole, it may be necessary to perform an analytic extension beyond the horizon, as in \cite{Herzog:2002pc}.

\section*{Acknowledgments}
We would like to thank J.L.F.~Barb\'on for useful comments and for suggesting the computation of the spectral function. We would also like to thank A.~Peet and D.T.~Son. This work was supported in part by the U.S. Department of Energy under
Grant No. DE-FG02-96ER40956.

%\newpage
%%%%%%%%%%%%%%%%%%%%%%%%%%%%%%%%%%%%
\appendix

%%%%%%%%%%%%%%%%%%%%%%%%%%%%%%%%
\section{Energy of $T>T_H$ black holes}\label{energycomp}

The energy of the black hole solutions can be evaluated using the results of
\cite{Hawking:1995fd}.
Consider a constant time hypersurface $\Sigma_t$ in the black hole background. The metric on
$\Sigma_t$ is given by
\begin{equation}
d s^2|_{\Sigma_t} = e^{\phi\over 2}\left[  R^2 d \Omega^2_3
+ \gamma {d r^2 \over f(r)} +N_c\alpha' d Y_5^2\right]\, , \quad
\gamma = \frac{N_c\alpha' R^2}{R^2 + N_c}
\label{metricSigmat}\end{equation}
Take $S^\infty_t$ to be a surface of constant and large radial coordinate value $r = r_{max}$ in $\Sigma_t$,
which we can think of as a boundary of $\Sigma_t$. Define $n_{bh}^\mu$ to be a unit vector normal to
$S^\infty_t$
$$
n_{bh}^\mu = \sqrt{f(r) \gamma^{-1}} e^{-\phi/4} \, \delta^\mu_r |_{r =r_{max}}
$$
The extrinsic curvature of $S^\infty_t$ in $\Sigma_t$ is
$$
K_{bh} = \nabla_\mu n^\mu_{bh} = \frac{1}{\sqrt{g_{\Sigma_t}}} \partial_\mu \left( \sqrt{g_{\Sigma_t}} n^\mu_{bh}   \right)
= \sqrt{f(r) \gamma^{-1}} e^{-9/4 \,\phi} \partial_r ( e^{2 \phi } )|_{r =r_{max}}
$$
where $g_{\Sigma_t}$ is the determinant of the metric on $\Sigma_t$ \eqref{metricSigmat}.
The energy of the black hole is defined as an integral over the boundary $S^\infty_t$ of
\begin{equation}
E = \lim_{r_{max} \to \infty} - \frac{1}{8 \pi G_{10}} \int K_{bh} \sqrt{| g_{tt} |} dS^\infty_t
\end{equation}
Using
$$
dS^\infty_t = e^{2 \phi} R^3 ( N_c \alpha' )^{5/2} d\Omega_3 dY_5 |_{r =r_{max}}
$$
the energy would be
$$
E
%= - \frac{1}{8 \pi G_{10}} \int \sqrt{f(r) \gamma^{-1}} e^{-9/4 \,\phi} \partial_r ( e^{2 \phi } )|_{r =r_{max}}
%\times \sqrt{f(r)} e^{\phi/4}
%\times e^{2 \phi}
%R^3 ( N_c \alpha' )^{5/2} d\Omega_3 dY_5 |_{r =r_{max}}
%$$
%$$
= - \frac{1}{8 \pi G_{10}} R^3 ( N_c \alpha' )^{5/2}  f(r) \gamma^{-1/2} \partial_r ( e^{2 \phi } )|_{r =r_{max}}  \int  d\Omega_3 dY_5
$$
The above expression is divergent and needs to be regularized. Using the background renormalization method, we
subtract the analogous contribution of a reference background
\begin{equation}
E_{ren} = - \frac{1}{8 \pi G_{10}} \int \left(  K_{bh} - K_{ref} \right) \sqrt{| g_{tt} |} dS^\infty_t
\end{equation}
For this procedure to be well-defined, the reference background needs to have the same asymptotics
as the black hole solution, and in particular its metric and fields need to approach each other sufficiently fast
as $r_{max} \to \infty$. The natural choice for these backgrounds is
\begin{equation}
d s^2 = e^{\phi\over 2}\left[ - dt^2 + R^2 d \Omega_3^2
+ \gamma {d r^2} +N_c\alpha' d Y_5^2\right]\,
\label{S3ref}\end{equation}
where the dilaton and three-form fields are exactly the same as for the black hole solutions \eqref{S3bhNNN}.
The radius $R$ of the three-sphere is also determined by the $3$-form flux \eqref{Rflux}.
The metrics \eqref{S3ref} are also solutions of the equations of motion of the type IIB action coupled to the flavor DBI action.

First of all, the Euclidean time direction of the reference solution has to be compactified and its radius at
$r = r_{max}$ has to match that of the black hole solution
$$
%e^{\phi_{bh}/2} f(r_{max}) \beta^2 = e^{\phi_{ref}/2} \beta'{}^2
|g_{tt,ref}(r_{max})| = e^{\phi_{bh}/2} f(r_{max}) \ .
$$
Then, for the metric and fields of the reference background to match those of the black hole solution we set
$$
\phi_{ref}(r) = \phi_{bh}(r) \ , \quad R_{ref} = R_{bh} \ .
$$

As we did above, we can evaluate the extrinsic curvature of a boundary surface $S^\infty_t$ in the reference background
$$
K_{ref} = \gamma^{-1/2} e^{-9/4 \,\phi} \partial_r ( e^{2 \phi } )|_{r =r_{max}}
$$
The renormalized energy becomes
$$
E_{ren} = - \frac{1}{8 \pi G_{10}} \int \left(  K_{bh} - K_{ref} \right) \sqrt{| g_{tt} |} dS^\infty_t
$$
$$
= - \lim_{r_{max} \to \infty} \, \frac{1}{8 \pi G_{10}} \gamma^{-1/2} R^3 ( N_c \alpha' )^{5/2}  \left[ f(r)^{1/2}  ( f(r)^{1/2} - 1 ) \partial_r ( e^{2 \phi } )|_{r =r_{max}}
\right] \int  d\Omega_3 dY_5
$$
\begin{equation}
=  \frac{1}{8 \pi G_{10}} e^{2 \phi_0 + 2 r_0} \gamma^{-1/2} R^3 ( N_c \alpha' )^{5/2} \int  d\Omega_3 dY_5
%=  \frac{1}{8 \pi G_{10}} e^{2 \phi_{horizon}} \gamma^{-1/2} R^3 ( N_c \alpha' )^{5/2} \int  d\Omega_3 dY_5
\end{equation}
where $\phi_0 + r_0$ is the value of the dilaton at the horizon radius $r_0$.

\noindent
The dependence of $E_{ren}$ on the temperature comes solely from the fact that the radius $R$ and $\gamma$ are
functions of $T$
$$
E_{ren} = \frac{1}{8 \pi G_{10}} e^{2 \phi_0 + 2 r_0} ( N_c \alpha' )^{5/2} \gamma^{-1/2}(T) R^3(T)  \int  d\Omega_3 dY_5
$$
\begin{equation}
= \frac{e^{2 \phi_0 + 2 r_0} (\alpha')^2 }{8 \pi G_{10}}  N_c^2 \, V(S^3) \, V(Y_5) {T\over T_H}  \, ,
\label{Eren}\end{equation}
where we have defined $V(S^3)=R^3\int d\Omega_3$ and $V(Y_5)=\int d Y_5$.

The energy of the black hole solutions is proportional to their horizon area
\begin{equation}
A = e^{2 \phi_0 + 2 r_0}  ( N_c \alpha' )^{5/2} \, V(S^3) \, V(Y_5) \ ,%R^3 \int d\Omega_3 dY_5
\end{equation}
and the free energy of the black holes is vanishing
$$
F = E_{ren} - T S = 0 \ ,
$$
where the entropy is
$$
S = \frac{A}{4 G_{10}} \ .
$$
The fact that the free energy vanishes can also be derived directly from a computation of the renormalized Euclidean action of the black holes ${\cal I}$, that is related to the free energy by  ${\cal I} = F/T$.

%%%%%%%%%%%%%%%%%%%%%%%%%%%%%%%%%%%%%%%%%%%%%%%%%%%%
\section{Equations of motion for metric and dilaton fluctuations}\label{sec:EOMcomputation}

The equations of motion for the metric and dilaton fluctuations around the
black hole background can be deduced from the following five-dimensional action
\begin{equation}
\int d^{5}x \sqrt{-g_5} \,
\left[ R_5 - \frac{4}{3} \left( \partial \phi \right)^2
+ \frac{4}{N_c}  \,e^{-4/3 \, \phi} \
\right]
\label{5dnewA}\end{equation}
The background black hole metric and dilaton are
\begin{align}
ds^2 = & \, e^{4/3\, \phi_b}
\left[ - ( 1 - u ) dt^2 + d\vec x_3^2 + N_c \alpha'
\, \frac{du^2}{4 u^2 ( 1 - u )}\, \right]
\nonumber \\
\phi_b = &  - \frac{1}{2} \log u
\label{bhgnewU}\end{align}
The equations of motion are
\begin{equation}
R_{\mu\nu} - \frac{1}{2} g_{\mu\nu} R - \frac{4}{3} \partial_\mu \phi \partial_\nu \phi
+ \frac{2}{3} ( \partial \phi )^2 g_{\mu\nu} - \, \frac{2}{N_c} \, e^{-4/3 \, \phi}\, g_{\mu\nu}
= 0
\label{eommetric}\end{equation}
\begin{equation}
\nabla^2 \phi - \frac{2}{N_c} \, e^{-4/3 \, \phi}\, = 0
\label{eomDila}
\end{equation}

\noindent
We will consider fluctuations $\delta g_{\mu\nu} = h_{\mu\nu}(u,t,z)$, $\delta \phi(u,t,z)$
in the form
\begin{align}
h_{tt} = & e^{- i \omega t + i q z} e^{4/3 \, \phi_b} (1-u) H_{tt}(u) \nonumber \\
h_{tz} = & e^{- i \omega t + i q z} e^{4/3 \, \phi_b} H_{tz}(u) \nonumber \\
h_{xx} = & \, h_{yy} = \frac{h_{aa}}{2}  \nonumber \\
h_{aa} = & e^{- i \omega t + i q z} e^{4/3 \, \phi_b} H_{aa}(u) \nonumber \\
h_{zz} = & e^{- i \omega t + i q z} e^{4/3 \, \phi_b} H_{zz}(u) \nonumber \\
\delta \phi = & e^{- i \omega t + i q z} \varphi(u) \nonumber \\
h_{tx} = & \, h_{ty} =  e^{- i \omega t + i q z} e^{4/3 \, \phi_b} H_{tx}(u) \nonumber \\
h_{xz} = & \, h_{yz} =  e^{- i \omega t + i q z} e^{4/3 \, \phi_b} H_{zx}(u) \nonumber \\
h_{xy} = &  e^{- i \omega t + i q z} e^{4/3 \, \phi_b} H_{xy}(u)
\end{align}
and the following choice of gauge
\begin{equation}
h_{tu} = h_{xu} = h_{yu} = h_{zu} = h_{uu} = 0 \ .
\label{gaugechoice}\end{equation}

\noindent
The sound mode metric and dilaton fluctuations $\{ H_{tt}, H_{tz}, H_{zz}, H_{aa}, \varphi \}$
satisfy the following system of linear differential equations
\begin{equation}
H_{tz}'' +  \frac{N_c q \omega}{4 u^2 (1 - u)} H_{aa} =  0
\nonumber
\end{equation}
\vspace{0.2cm}
\begin{equation}
- 12 u ( 1 - u ) ( H_{aa}'' + H_{zz}'' ) + 6 u ( H_{aa}' + H_{zz}' ) + 16 ( 1 - u ) \varphi'
+ \frac{3 N_c q^2}{u} H_{aa} - \frac{16}{u} \varphi = 0
\nonumber \\
\end{equation}
\vspace{0.2cm}
$$
12 u^2 ( 1 - u )^2 ( H_{aa}'' - 2 H_{tt}'' + 2 H_{zz}'' )
- 4 u ( 1 - u ) ( 3 u H_{aa}' - 9 u H_{tt}' + 6 u H_{zz}' + 8(1-u) \varphi' )
$$
\begin{equation}
+ 3 N_c ( \omega^2 - q^2 (1-u) ) H_{aa} + 6 N_c q^2 (1-u) H_{tt} + 12 N_c q \omega H_{tz}
+ 6 N_c \omega^2 H_{zz} + 32(1-u) \varphi =  0
\nonumber \\
\end{equation}
\vspace{0.2cm}
\begin{equation}
12 ( 1 - u )^2 ( H_{aa}'' - H_{tt}'' ) - 12 ( 1 - u ) H_{aa}' + 18(1-u) H_{tt}'
- \frac{16(1-u)^2}{u} \varphi'
\nonumber
\end{equation}
$$
+ \frac{3 N_c \omega^2}{u^2} H_{aa} + \frac{16(1-u)}{u^2} \varphi
= 0
$$
\vspace{0.2cm}
\begin{equation}
24 (1-u)^2 \varphi'' - 24 (1-u) \varphi' - \frac{6(1-u)^2}{u} ( \, H_{aa}' - H_{tt}' + H_{zz}'\, )
+ \frac{2}{u^2} \left( \, 8(1-u) + 3 N_c ( \omega^2 - (1-u) q^2 ) \, \right) \varphi = 0
\nonumber
\end{equation}
\vspace{0.2cm}
\begin{equation}
6 \omega ( 1 - u ) ( H_{aa}' + H_{zz}' ) + 6 q ( 1 - u ) H_{tz}'
+  3 \omega ( H_{aa} + H_{zz} ) + 6 q H_{tz} - 8 \frac{\omega (1 - u)}{u} \varphi =  0
\nonumber
\end{equation}
\vspace{0.2cm}
$$
6 q (1-u) ( H_{aa}' - H_{tt}' ) - 6 \omega H_{tz}' + 3 q H_{tt} - \frac{8 q (1-u)}{u} \varphi = 0
$$
\vspace{0.2cm}
$$
2 u ( 1 - u ) \left( \ 3 ( u - 2 ) ( H_{aa}' + H_{zz}' )  + 6 ( 1 - u ) H_{tt}' + 8(1-u) \varphi' \
\right)
$$
\begin{equation}
+ 3 N_c ( \omega^2 - (1-u) q^2 ) H_{aa} + 3 N_c q^2 (1-u) H_{tt} + 6 N_c q \omega H_{tz} +
3 N_c \omega^2 H_{zz}  + 16(1-u) \varphi = 0
\label{soundsystem}\end{equation}

\noindent
Conversely, the shear mode fluctuations $\{ H_{tx}, H_{zx} \}$ satisfy the system
$$
\omega H_{tx}' + q (1-u) H_{zx}' = 0 \ ,
$$
$$
H_{tx}'' - \frac{N_c}{4(1-u)u^2} ( \omega q H_{zx} + q^2 H_{tx} ) = 0 \ ,
$$
\begin{equation}
H_{zx}'' - \frac{1}{1-u} H_{zx}' + \frac{N_c}{4u^2(1-u)^2} ( \omega^2 H_{zx} + \omega q H_{tx} ) = 0 \ .
\label{shearsystem}\end{equation}

\noindent
Finally, the scalar mode fluctuation $H_{xy}$ satisfies the equation
\begin{equation}
H_{xy}'' - \frac{1}{1-u} H_{xy}' + N_c \frac{\omega^2 - q^2(1-u)}{4 u^2 (1-u)^2} H_{xy} = 0 \ .
\label{scalarequation}\end{equation}

\noindent
The functions $\{ H_{tt}, H_{tz}, H_{zz}, H_{tx}, H_{zx}, H_{xy}, \varphi \}$ are in general not invariant under
the residual diffeomorphisms that preserve the gauge choice \eqref{gaugechoice}
$$
x^\mu \rightarrow x^\mu + \xi^\mu \ ,
$$
$$
h_{\mu\nu} \rightarrow h_{\mu\nu} - \nabla_\mu \xi_\nu - \nabla_\nu \xi_\mu \ ,
$$
$$
\phi \rightarrow \phi - \xi^\mu \partial_\mu \phi \ .
$$
It is convenient to introduce the following gauge invariant variables \cite{qnms}
\begin{equation}
Z _{sound} = q^2 H_{tt} + 2 q \omega H_{tz} + \omega^2 H_{zz} + q^2 ( 1 - u )
\left( 1 + \frac{3 u}{2(1-u)} - \frac{\omega^2}{q^2(1-u)} \right) \frac{H_{aa}}{2} \ ,
\end{equation}
\begin{equation}
Z_\phi = \varphi - \frac{3}{8} H_{aa} \ ,
\end{equation}
\begin{equation}
Z_{shear} = q H_{tx} + \omega H_{zx}  \ ,
\end{equation}
\begin{equation}
Z_{scalar} = H_{xy}  \ ,
\end{equation}
The differential systems \eqref{soundsystem}\eqref{shearsystem}\eqref{scalarequation} then lead to
\begin{equation}
Z_\phi'' - \frac{1}{1-u} Z_\phi' + \frac{N_c \left( \omega^2 - (1-u) q^2 \right)}{4 u^2 (1-u)^2} Z_\phi = 0
\label{EQdilaton}\end{equation}
$$
Z_{sound}'' + \frac{q^2 u - 2 \omega^2}{(1-u)(2 \omega^2 + q^2(u-2))}\, Z_{sound}'
- \frac{N_c (4 q^4 - 8 q^2 \omega^2)}{3(1-u)(2 \omega^2 + q^2(u-2))}\, Z_\phi
$$
\begin{equation}
- \frac{4 q^2 u^2(1-u) - N_c ( q^4 (u-2)(u-1) + q^2 \omega^2 (3u-4) + 2 \omega^4 )}{4 u^2(1-u)^2(2 \omega^2 + q^2(u-2))}\, Z_{sound} = 0 \ ,
\label{EQscalarmetric}\end{equation}
\begin{equation}
Z_{shear}'' - \frac{\omega^2}{(1-u)(\omega^2 - q^2(1-u))} Z_{shear}' + \frac{N_c (\omega^2 - (1-u) q^2 )}{4 u^2(1-u)^2} Z_{shear} = 0 \ .
\end{equation}
\begin{equation}
Z_{scalar}'' - \frac{1}{1-u} Z_{scalar}' + \frac{N_c (\omega^2 - (1-u) q^2 )}{4 u^2 (1-u)^2} Z_{scalar} = 0 \ .
\end{equation}
Note that $Z_\phi$ and $Z_{scalar}$ satisfy the same differential equation, which corresponds to the equation of a minimally coupled scalar
in the black hole background.

%%%%%%%%%%%%%%%%%%%%%%%%%%%%%%%%%%%%%%%%%%%%%%%%%%%%%%%%%%%%%%%%%%%%%%%%%%
%%%%%%%%%%%%%%%%%                                      %%%%%%%%%%%%%%%%%%%
%%%%%%%%%%%%%%%%%             BIBLIOGRAPHY             %%%%%%%%%%%%%%%%%%%
%%%%%%%%%%%%%%%%%                                      %%%%%%%%%%%%%%%%%%%
%%%%%%%%%%%%%%%%%%%%%%%%%%%%%%%%%%%%%%%%%%%%%%%%%%%%%%%%%%%%%%%%%%%%%%%%%%

\end{document}